\documentclass[11pt,a4paper]{article}
\usepackage{jheppub}
\usepackage{amsmath,amssymb}
\usepackage{slashed}
\usepackage{graphicx}%
\usepackage{subfigure}
\usepackage{mathptmx}
\usepackage{url}
\newcommand{\beq}{\begin{equation}}
\newcommand{\eeq}{\end{equation}}

\newcommand{\be}{\begin{eqnarray}}
\newcommand{\ee}{\end{eqnarray}}
\long\def\hidestart#1\hideend{}
\title
{Pion and nucleon in two flavour QCD with unimproved Wilson fermions}
\author{Abhishek Chowdhury$^{a}$,}
\author{Asit K. De$^{a}$,}
\author{Sangita De Sarkar$^{a}$,}
\author{A. Harindranath$^{a}$,}
\author{Jyotirmoy Maiti$^{b}$,}
\author{Santanu Mondal$^{a}$ and}
\author{Anwesa Sarkar$^{a}$}

\affiliation{$^{a}$Theory Division, Saha Institute of Nuclear Physics \\
 1/AF Bidhan Nagar, Kolkata 700064, India}

\affiliation{$^{b}$Department of Physics, Barasat Government College,\\
10 KNC Road, Barasat, Kolkata 700124, India}

\emailAdd{abhishek.chowdhury@saha.ac.in}
\emailAdd{asitk.de@saha.ac.in}
\emailAdd{sangita.desarkar@saha.ac.in}
\emailAdd{a.harindranath@saha.ac.in}
\emailAdd{jyotirmoy.maiti@gmail.com}
\emailAdd{santanu.mondal@saha.ac.in}
\emailAdd{anwesa.sarkar@saha.ac.in}

\date{October 11, 2012}
\abstract{We calculate pion mass, pion decay constant, PCAC quark mass and nucleon mass
in two flavour lattice QCD with unimproved Wilson fermion and gauge actions. 
Simulations are performed using DD-HMC algorithm at two lattice spacings and two volumes 
for several values of the quark mass. The cutoff effects in pion mass and nucleon mass
for the explored region of parameter space are found to be negligible. The chiral behaviours of 
pion mass, pion decay constant and quark condensate are found to be qualitatively consistent with
NLO chiral perturbation theory. }
\begin{document}
\maketitle

\section{Introduction}

Because of the explicit violation  of chiral symmetry by a dimension
five kinetic operator, Wilson formulation has been known to be 
difficult to simulate at  
light quark masses. Lack of chiral symmetry means that the $``$physical''
quark mass is no longer proportional to the bare quark mass (the quark mass
renormalization is no longer only multiplicative) and
Wilson-Dirac operator is
not protected from arbitrarily small eigenvalues and may lead to zero or near
zero modes for individual configurations. This is the infamous problem of 
$``$exceptional configurations''. This leads to convergence difficulties for
fermion matrix inversion. This poses difficulties for
lattice simulations with Wilson fermions in the chiral region.

In the past, simulations with unimproved 
Wilson action has shown large scaling violations in hardonic observables. 
However, one should keep in mind that most of these were quenched simulations 
done at large pion masses, not small enough lattice spacings and smaller 
volumes. 
Further, the demonstration of the suppression of topological susceptibility
with decreasing quark mass was inconclusive. The chiral behaviour of pion 
mass and decay constant with respect to quark mass (specifically, the 
presence of chiral logarithms) as dictated by chiral perturbation theory 
was also not convincingly demonstrable in the past with un-improved Wilson 
fermions. All these issues raise the question: Does Wilson lattice QCD 
belong to the same universality class as continuum QCD?      

The situation regarding $``$exceptional configurations'' has improved  
partly due to the
finding \cite{deldebbiojhep} employing DD-HMC algorithm \cite{ddhmc}
that the numerical simulations are safe
from accidental zero modes for large volumes. Simulations with unimproved Wilson 
fermions at smaller quark masses and lattice spacings and larger volumes have
become possible with the DD-HMC algorithm. 
As part of an on-going 
program \cite{p1,p2} to study the chiral properties of
Wilson lattice QCD (unimproved fermion and gauge actions), recently, 
we have demonstrated the suppression of 
topological susceptibility with decreasing quark mass in the case of
unimproved Wilson fermion and gauge action \cite{ac1,ac2} where,    
the suppression of topological susceptibility
 with decreasing volume was also shown.    
In order to shed 
light on the mechanisms leading to these suppressions, we have further 
carried out a detailed study of the  two-point topological charge density
correlator \cite{ac3}. An exploratory investigation of the autocorrelations
of various observables with DD-HMC algorithm is presented in Ref.
\cite{ac4}. In these works we have employed ensembles of gauge configurations generated by means of   
DD-HMC \cite{ddhmc}
algorithm using unimproved Wilson fermion and Wilson gauge  
actions \cite{wilson} with 
$n_f=2$ mass degenerate quark flavours.

In this work we investigate pion mass ($m_\pi$) and decay constant, PCAC quark mass, quark condensate and nucleon mass
in the range $290\lesssim m_\pi\lesssim750$ MeV. We perform  qualitative chiral extrapolations
of various observables in the range  $350\lesssim  m_\pi\lesssim550$ MeV.
So far the simulations are done 
at two lattice spacings 
in the region of 0.05 - 0.07 fm where
many of the modern simulations of LQCD with improved actions are carried out.

\section{Simulation and Observables}
Simulations have been carried out at two values of the gauge coupling correspond to
$\beta$=5.6 and 5.8.
 At $\beta=5.6$ the lattice volumes are 
 $24^3 \times 48$ and $32^3 \times 64$ and at $\beta=5.8$ the lattice volume
 is $32^3 \times 64$. The number of thermalized gauge configurations ranges 
from $3760$ to $13646$.
The lattice parameters and simulation statistics are given in Table 
\ref{table1}. For all ensembles of configurations the average 
Metropolis acceptance rates range between $75-98\%$.
For pion and nucleon 
we consider the following zero spatial momentum correlation
functions
\begin{eqnarray}
& & C(t)~~ =~~\langle 0 \mid {\cal O}_1(t) {\cal O}_2(0)
\mid 0 \rangle~ 
\end{eqnarray}
where $t$ refers to Euclidean time.
For the nucleon 
${\cal O}_1 {\cal O}_2\equiv{\rm N} {\overline {\rm N}}$ with ${\rm N}= (q_d^T C \gamma_5 q_u) q_u$.  
For the pion ${\cal O}_1 {\cal O}_2\equiv P P^\dagger$,  $A A^\dagger$, 
 $A P^\dagger$ or $P A^\dagger$
where  $P = {\overline q}_i \gamma_5 q_j$ (pseudoscalar
density)  and $A$ corresponds to
$A_4 = {\overline q}_i \gamma_4 \gamma_5 q_j$ (fourth component of the 
axial vector current). Here $i$ and $j$ stand for
flavor indices for the $u$ and $d$ quarks and  for the charged pion $i\ne
j$. For pion we use point source and point sink and 
for nucleon we use wall source and point sink.
Unless otherwise stated  $20$ HYP smearing steps
with optimized smearing coefficients $\alpha_1 =0.75$,
$\alpha_2=0.6$ and $\alpha_3=0.3$ \cite{hasenfratz2} are used 
for the gauge observables. 
\begin{table}
\begin{center}
\begin{tabular}{|l|l|l|l|l|l|l|l|l|l|l|}

 \multicolumn{8}{c}{$\beta = 5.6$} \\
\hline
$tag$&$lattice$& $\kappa$& $block $&{$N_2$}& {$N_{cfg}$}  &{$\tau$}&$r_0 \times m_{\pi}$\\ 
\hline

{$B_{1b}$}&{$24^3\times48$}&{$0.1575$}&{$12^2\times6^2$}&{$18$}&{$13128$}&{$0.5$}&$1.7719(38)$ \\
{$B_{3b}$}&{$~~~~~,,$}&{$0.158$}&{$12^2\times6^2$}&{$18$}&{$13646$}&{$0.5$}&$1.2542(58)$ \\
{$B_{4b}$}&{$~~~~~,,$}&{$0.158125$}&{$12^2\times6^2$}&{$18$}&{$11328$}&{$0.5$}&$1.0925(58)$\\

\hline
{$C_1$}&{$32^3\times64$}&{$0.15775$}&{$8^3\times16$}&{$8$}&{$6844$}&{$0.5$}&$1.5345(54)$\\
{$C_2$}&{$~~~~~,,$}&{$0.158$}&{$8^3\times16$}&{$8$}&{$7576$}&{$0.5$}&$1.2590(59)$\\
{$C_3$}&{$~~~~~,,$}&{$0.158125$}&{$8^3\times16$}&{$8$}&{$8768$}&{$0.5$}&$1.1010(60)$\\
{$C_4$}&{$~~~~~,,$}&{$0.15815$}&{$8^3\times16$}&{$8$}&{$9556$}&{$0.5$}&$1.0697(57)$\\
{$C_5$}&{$~~~~~,,$}&{$0.15825$}&{$8^3\times16$}&{$8$}&{$11520$}&{$0.5$}&$0.9343(55)$\\
{$C_6$}&{$~~~~~,,$}&{$0.1583$}&{$8^3\times16$}&{$8$}&{$4384$}&{$0.25$}&$0.8476(99)$\\

\hline \hline

  \multicolumn{8}{c}{$\beta = 5.8$} \\
\hline
$tag$&$lattice$& $\kappa$& $block$ &{$N_2$}& {$N_{cfg}$}  &{$\tau$}& $r_0 \times m_{\pi}$\\
\hline
{$D_{1a}$}&{$32^3\times64$}&{$0.1543$}&{$8^3\times16$}&{$8$}&{$9600$}&{$0.5$}&$1.3259(76)$\\
{$D_{2b}$}&{$~~~~~,,$}&{$0.15445$}&{$8^3\times16$}&{$24$}&{$4800$}&{$0.5$}&$1.1138(73)$\\
{$D_{3a}$}&{$~~~~~,,$}&{$0.15455$}&{$8^3\times16$}&{$8$}&{$12160$}&{$0.5$}&$0.9968(87)$\\
{$D_{4b}$}&{$~~~~~,,$}&{$0.15462$}&{$8^3\times16$}&{$24$}&{$7528$}&{$0.5$}&$0.8637(81)$\\
{$D_{5b}$}&{$~~~~~,,$}&{$0.15466$}&{$8^3\times16$}&{$24$}&{$3760$}&{$0.5$}&$0.8360(131)$\\
{$D_{6b}$}&{$~~~~~,,$}&{$0.1547$}&{$8^3\times16$}&{$24$}&{$4256$}&{$0.5$}&$0.6851(181)$\\
\hline \hline
\end{tabular}
\end{center}
\caption{Lattice parameters, simulation statistics and pion mass ($m_{\pi}$) in the unit of Sommer parameter ( $r_0$).
Here $block$, $N_2$, $N_{cfg}$, $\tau$
refers to DD-HMC block, step number for the force $F_2$, number of 
DD-HMC configurations and the Molecular Dynamics trajectory length respectively.   }
\label{table1}
\end{table}

The pion decay constant $F_\pi$ and the quark mass $m_q$ from PCAC or 
the axial Ward identity (AWI) are respectively defined, in the
continuum as follows:
\begin{eqnarray}
\langle 0\mid A_\mu(0)\mid \pi(p)\rangle &=&  
F_\pi p_\mu, \label{fpi}\\
\partial_\mu A_\mu (x) &=& 2 m_q P(x). \label{PCAC}
\end {eqnarray}
From the PP and the AP propagators
\begin{eqnarray}
C^{PP} &=&  \frac{1}{2 m_\pi} \mid \langle 0 \mid P(0) 
\mid \pi \rangle \mid^2 ~,\\
C^{AP} &=& \frac{1}{2 m_\pi}  \langle 0 \mid A_4(0) \mid \pi \rangle
\langle \pi \mid P^{\dagger}(0) \mid 0 \rangle
\end{eqnarray}
which lead to
\begin{eqnarray}
 F_\pi^{AP} ~ = ~ \frac{2\kappa ~ C^{AP}}{\sqrt{ m_\pi C^{PP}}} \label{fpiAP}.
 \end{eqnarray}
Using PCAC
\begin{eqnarray}
\partial_\mu \langle 0 \mid A_\mu (x)P^\dagger(0) \mid 0 \rangle =
2 m_q \langle 0 \mid P (x)P^\dagger(0) \mid 0 \rangle~.
\end{eqnarray}
Summing over spatial coordinates 
\begin{eqnarray}
\sum_{\bf x} \partial_\mu \langle 0 \mid 
A_\mu (x)P^\dagger(0) \mid 0 \rangle =
2 m_q \sum_{\bf x} 
\langle 0 \mid P (x)P^\dagger(0) \mid 0 \rangle~.
\end{eqnarray}
At large $t$,
\begin{eqnarray}
\partial_4 C^{AP}\Big [ e^{-m_\pi t} - e^{-m_\pi (T-t)}\Big ] = 2 m_q C^{PP}
\Big [ e^{-m_\pi t} + e^{-m_\pi (T-t)}\Big ]
\end{eqnarray}
which leads to 
\begin{eqnarray}
 m_q^{AP} = \frac{m_\pi}{2} ~ \frac{C^{AP}}{C^{PP}}~. \label{mqap}
\end{eqnarray}
\section{Computation of pion mass and decay constant, PCAC quark mass and nucleon mass }
\begin{figure}
 \includegraphics[width=3.5in,clip]
{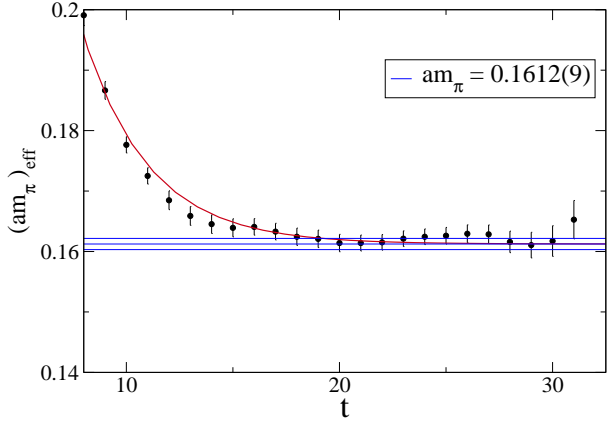}
\caption{Effective mass $am_\pi$ versus $t$ for $\beta=5.8$, $\kappa=0.1543$ and the volume $32^3\times 64$.}
\label{ampi}
\end{figure}
\begin{figure}
\subfigure{
 \includegraphics[width=3.5in,clip]
{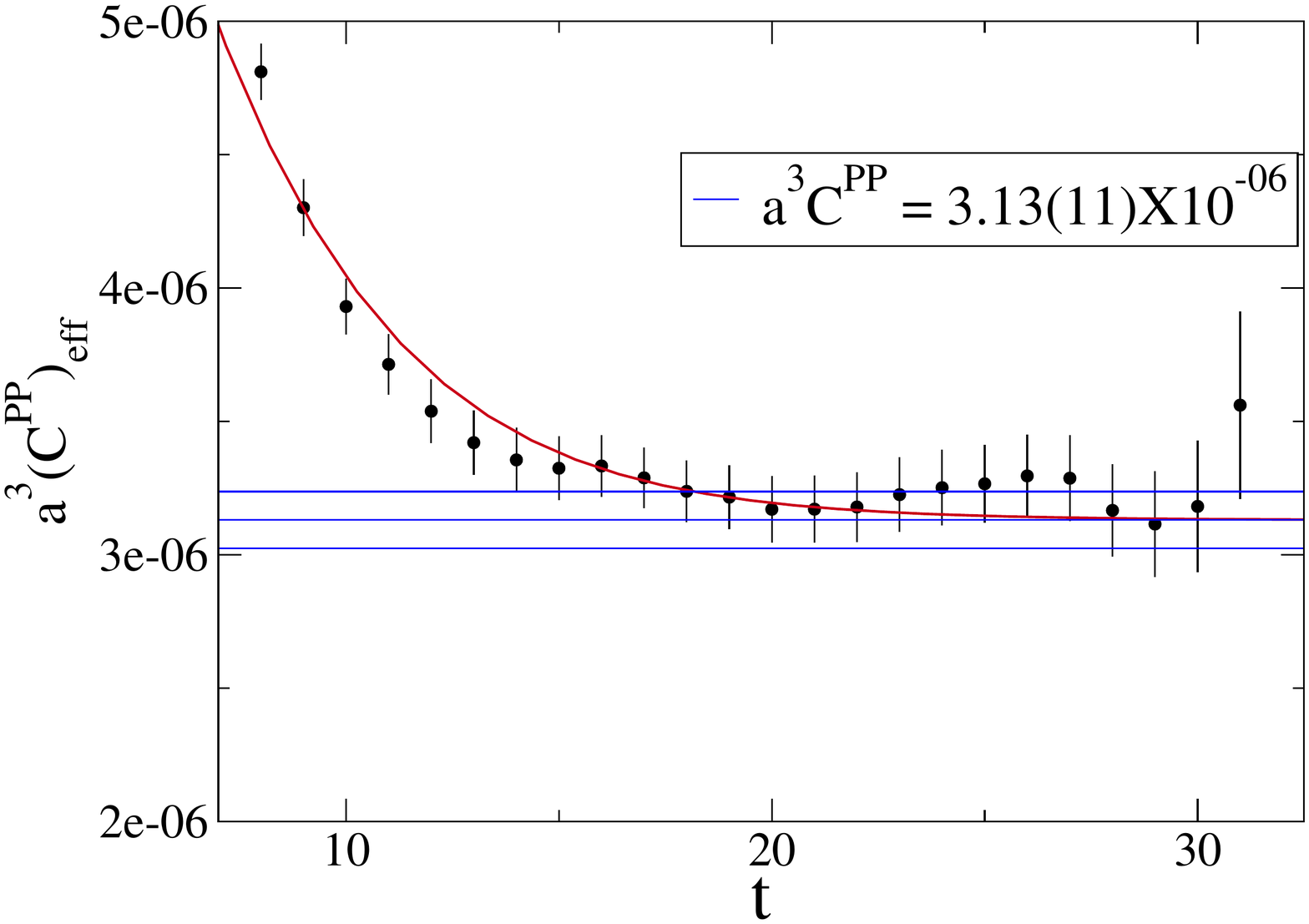}}
\subfigure{
 \includegraphics[width=3.5in,clip]
{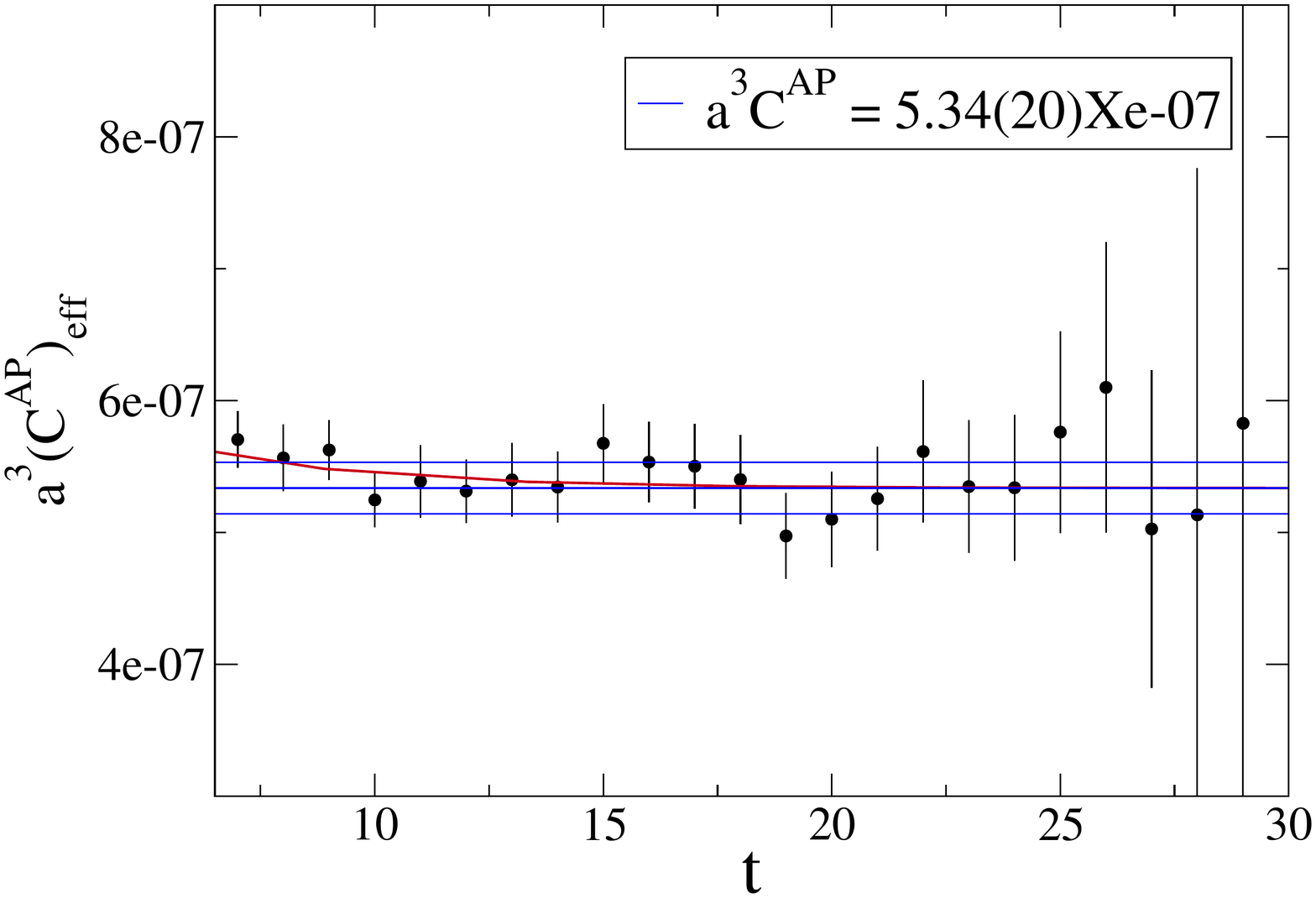}}
\caption{Effective coefficients  $a^3C_{PP}$ (left) and $a^3C_{AP}$ (right) versus 
$t$ for $\beta=5.8$, $\kappa=0.1543$ and  the volume $32^3\times 64$.}
\label{cppcap}
\end{figure}

\begin{figure}
\subfigure{
 \includegraphics[width=3.5in,clip]
{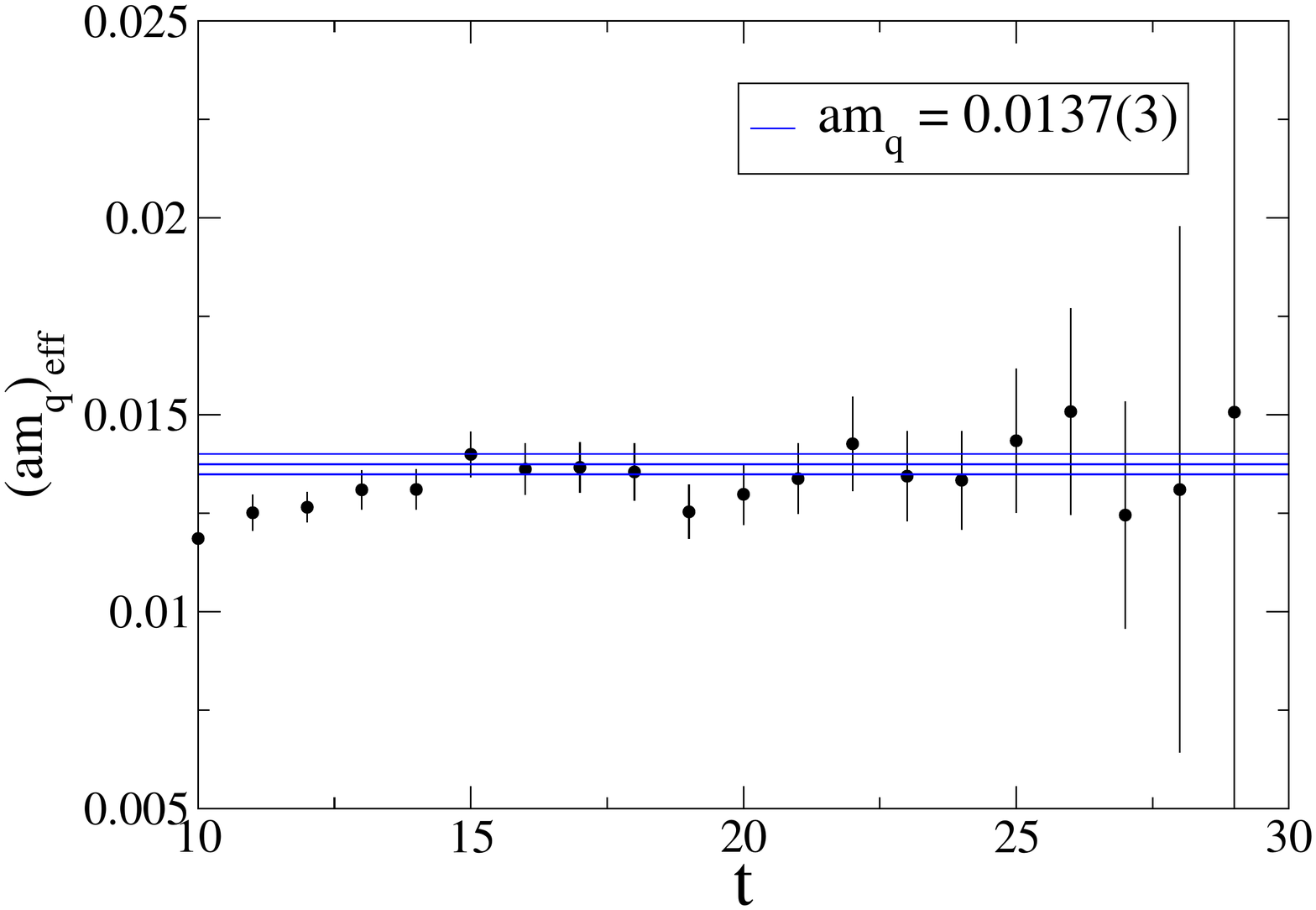}}
\subfigure{
 \includegraphics[width=3.5in,clip]
{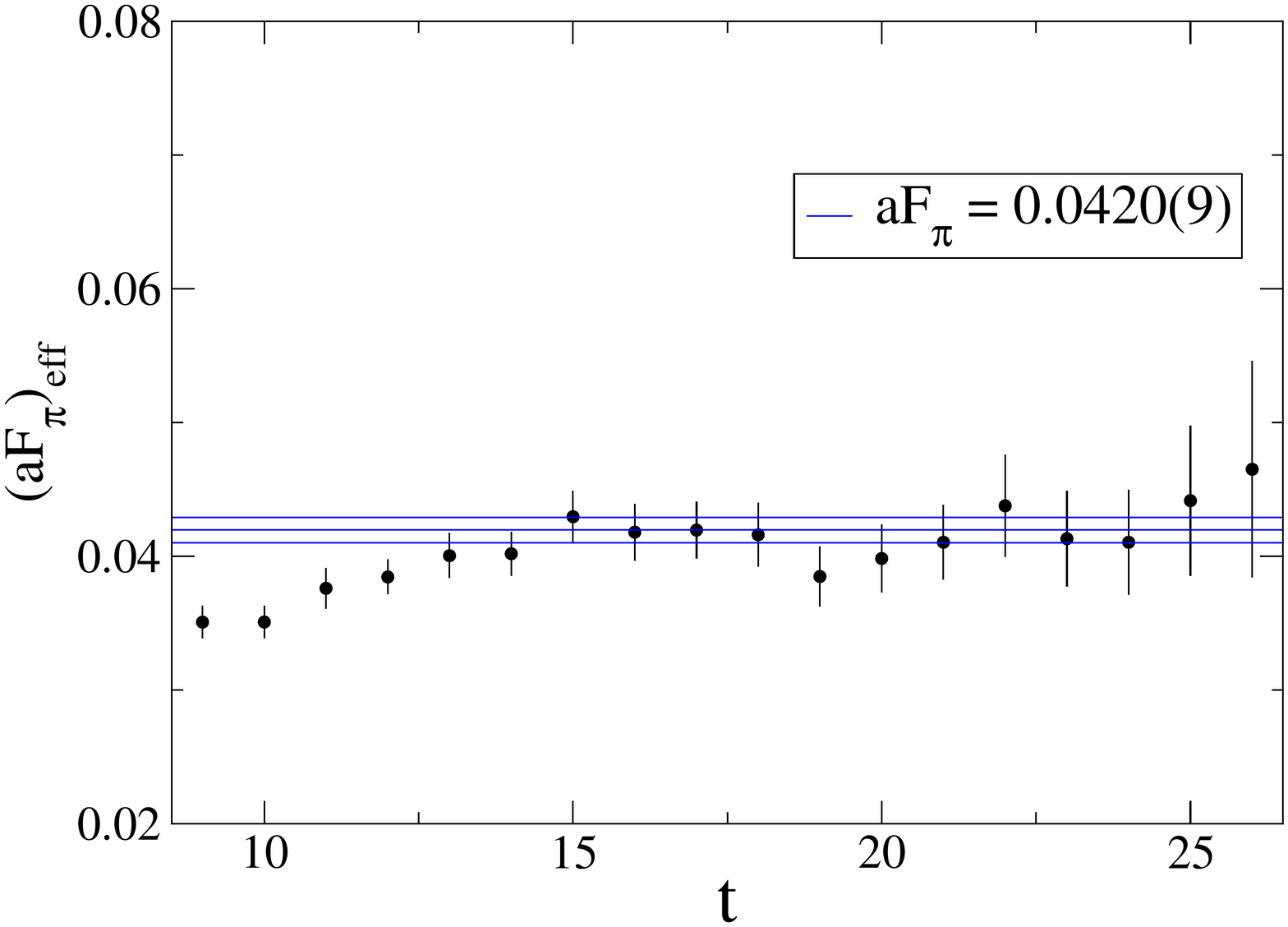}}
\caption{Effective $am_q^{AP}$ and effective  $aF_\pi^{AP}$ versus $t$ for 
for $\beta=5.8$, $\kappa=0.1543$ and  the volume $32^3\times 64$.}
\label{mqfpi}
\end{figure}

The two point correlation function $C(t)$ (with point source and point sink) may be expanded as
\beq
C(t)= c_0~ e^{-M_0t}~+~c_1~ e^{-M_1t}+~....
\eeq
where $M_0$ denotes the mass of the ground state and $M_1$ is the mass of
the first excited state.
The effective mass and the effective coefficient are given by
\be
M_{eff}(t)&=&M_0\left[1+\frac{2c_1}{c_0}e^{-(M_1-M_0)t}\right]\label{me}\\
C_{eff}(t)&=&c_0\left[1+\frac{c_1}{c_0}(1+2 M_0 t)e^{-2M_0t}\right]\label{ce}
\ee
We first calculate effective pion mass ($a(m_\pi)_{eff}$) from both $PP$ and $AP$ correlators. Then
Eq. (\ref{me}) is used to determine the asymptotic value $aM_0$ of $aM_{eff}(t)$.
Similarly we determine the asymptotic values of $a^3C^{PP}$ and $a^3C^{AP}$ by fitting 
$a^3C^{PP}_{eff}$ and 
$a^3C^{AP}_{eff}$ respectively with the Eq. (\ref{ce}).

The sample plots illustrating the dependence of these observables on $t$ are
given in figures \ref{ampi} and \ref{cppcap} for $\beta=5.8$, $\kappa=0.1543$ 
and the volume $32^3\times 64$. The fitting ranges in t are 8-31 and 8-29 for $PP$ and 
$AP$ correlators respectively. 
 In these figures the horizontal line with error
bars represent the asymptotic value giving  the mass and coefficients.  

The effective quark mass and the effective pion decay constant are calculated 
by using Eq. (\ref{mqap}) and Eq. (\ref{fpiAP}) respectively.
The sample plots illustrating the dependence of $(am_q^{AP})_{eff}$ and $(aF_\pi^{AP})_{eff}$
on $t$ are given in figure \ref{mqfpi} for $\beta=5.8$, $\kappa=0.1543$ 
and the volume $32^3\times 64$. In these figures the horizontal line with error
bars represent the quark mass $am_q^{AP}$ and pion decay constant $aF_\pi^{AP}$ calculated using
the asymptotic values of  the mass and coefficients from Eq. (\ref{mqap}) and Eq. (\ref{fpiAP}).

\begin{figure}
\subfigure{
 \includegraphics[width=3.5in,clip]
{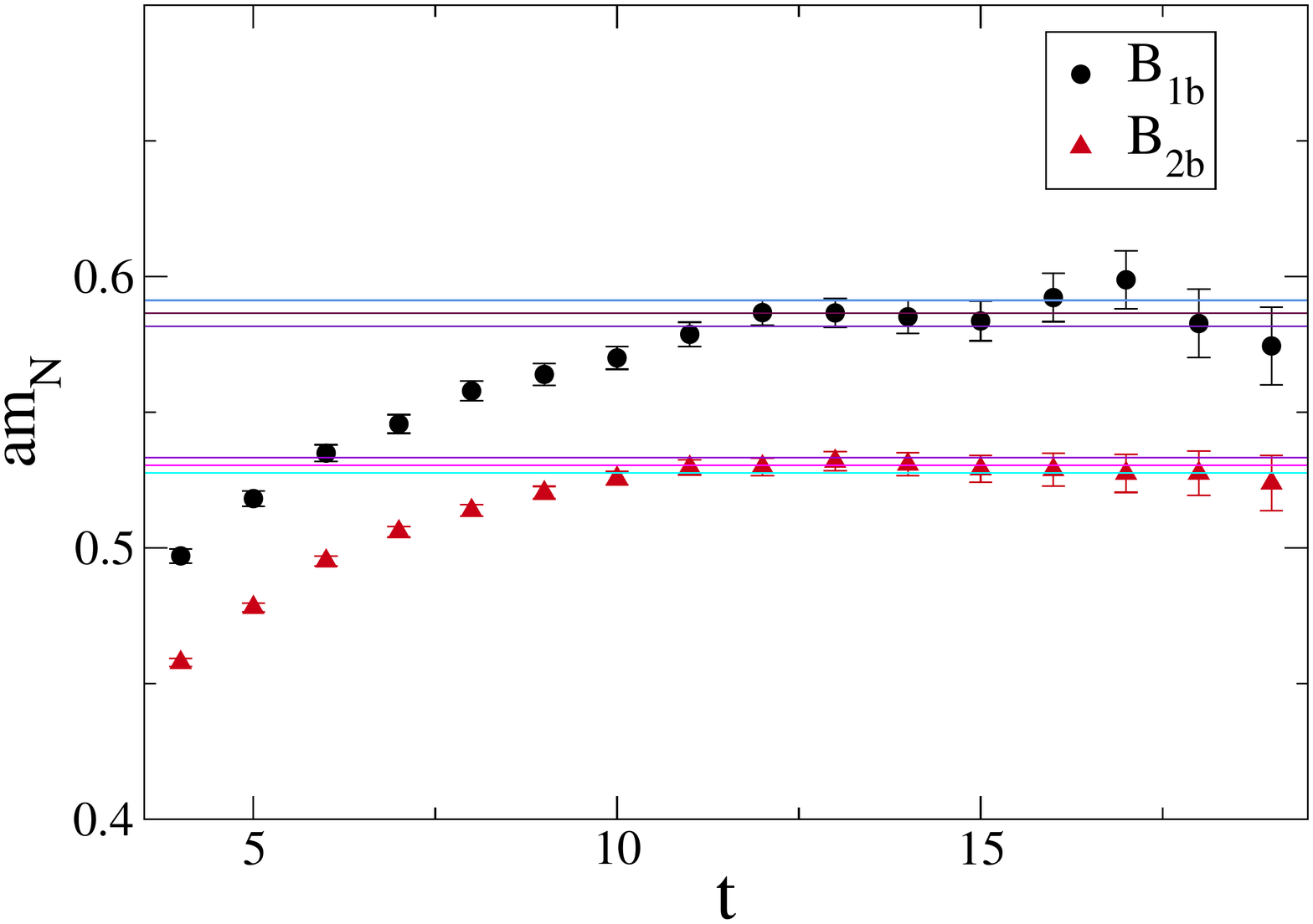}}
\subfigure{
 \includegraphics[width=3.5in,clip]
{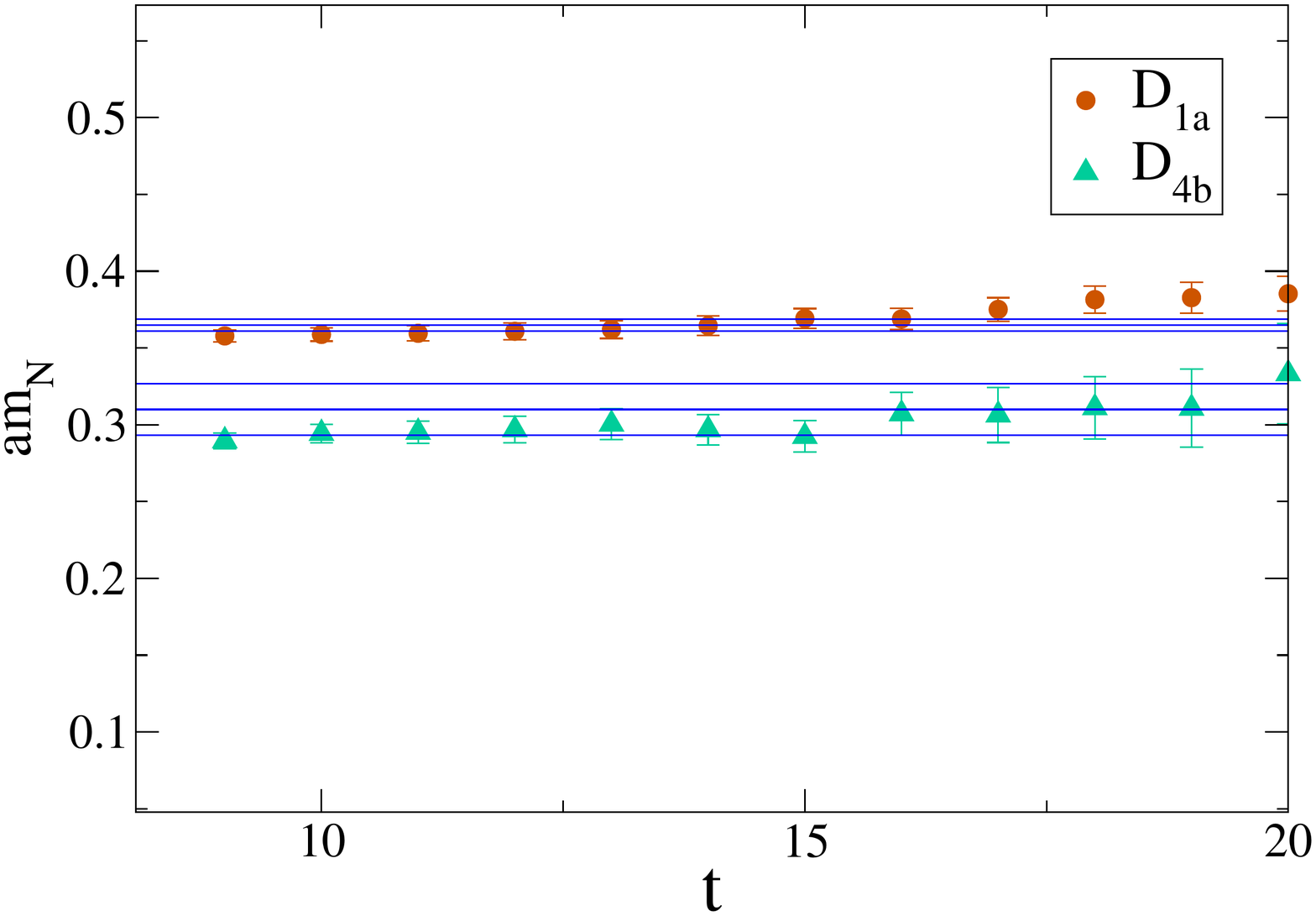}}
\caption{Typical plots of nucleon effective mass at $\beta$=5.6 (left) and
 $\beta$=5.8 (right).}
\label{mn}
\end{figure}

We have extracted nucleon mass using wall source.
Plots of the effective mass of nucleon at $\beta$=5.6, lattice volume 
$24^3 \times 48$ and $\kappa$= 0.1575, 0.15775 are shown in the Fig. \ref{mn}
(left) where the fitting ranges are t = 11-16 and 11-14 respectively.
Similarly in Fig. \ref{mn} (right) we have shown the effective mass 
of nucleon at $\beta$=5.8, lattice volume 
$32^3 \times 64$ for $\kappa$= 0.1543 and 0.15462 with fitting ranges t = 9-25 and 16-20 respectively.
The effective masses are extracted from the linear fit to the plateau region. 
In table \ref{table2} we present the lattice data for $am_{\pi}$,  $am_q$, $aF_{\pi}$ and $am_N$.
At $\kappa=0.15466$ for $\beta=5.8$, the signal for effective mass of the nucleon was noisy so we
do not quote a number.   

From table \ref{table2} it is clear that finite volume effect is negligible for 
the pion mass and nucleon mass at $\beta$=5.6 and lattice volume $24^3\times 48$.
At $\beta$=5.6 and lattice volume $32^3\times 64$ for lightest quark masses studied 
$m_\pi L$ values are 4.67 and 4.23 for $\kappa$ = 0.15825 and 0.1583 respectively. Note that the 
physical volume at $\beta$=5.6 and lattice volume $24^3\times 48$ is very
close to the physical volume at $\beta$=5.8 and lattice 
volume $32^3\times 64$. Thus comparing the values of $r_0 m_\pi$ from table \ref{table1}
for $\beta$=5.8 we expect that finite volume effect should be negligible upto and including $\kappa=
0.15455$ at lattice volume $32^3\times 64$. At $\beta$=5.8 for $\kappa = 0.15462$ and 0.15466 
the $m_\pi L$ values are 3.36 and 3.25 respectively. Only for 
the smallest quark mass ($\kappa=0.1547$) $m_\pi L =2.67$. Hence qualitatively we expect that finite
volume effect would be negligible for pion mass and nucleon mass for the region of parameter space 
studied except possibly for the lightest quark mass at $\beta$=5.8. According
to NLO Chiral Perturbation Theory ($\chi$PT), the finite volume effect for $F_\pi$ is four times that of $m_\pi$. Thus at 
$\beta$=5.8 for $\kappa$ = 0.15462 and higher we can expect non negligible finite volume effect for
$F_\pi$.

\begin{table}
\begin{center}
\begin{tabular}{|l|l|l|l|l|l|}

 \multicolumn{6}{c}{$\beta = 5.6$} \\
\hline
$lattice$&$\kappa$& $am_{\pi}$& $am_q$&{$aF_{\pi}$}& {$am_N$}\\ 
\hline

{$24^3\times48$}&{$0.1575$}&{$0.2764(6)$}&{$0.0277(3)$}&{$0.0655(7)$}&{$0.5865(48)$} \\
{$~~~~~,,$}&{$0.158$}&{$0.1957(9)$}&{$0.0148(3)$}&{$0.0549(12)$}&{$0.4735(65)$} \\
{$~~~~~,,$}&{$0.158125$}&{$0.1704(9)$}&{$0.0110(3)$}&{$0.0502(13)$}&{$0.4450(18)$}\\

\hline
{$32^3\times64$}&{$0.15775$}&{$0.2394(8)$}&{$0.0211(2)$}&{$0.0599(9)$}&{$0.5304(28)$}\\
{$~~~~~,,$}&{$0.158$}&{$0.1964(9)$}&{$0.0145(2)$}&{$0.0543(10)$}&{$0.4676(51)$}\\
{$~~~~~,,$}&{$0.158125$}&{$0.1718(9)$}&{$0.0111(2)$}&{$0.0503(10)$}&{$0.4389(134)$}\\
{$~~~~~,,$}&{$0.15815$}&{$0.1669(9)$}&{$0.0110(3)$}&{$0.0511(15)$}&{$0.4425(64)$}\\
{$~~~~~,,$}&{$0.15825$}&{$0.1458(9)$}&{$0.0078(1)$}&{$0.0474(9)$}&{$0.4045(61)$}\\
{$~~~~~,,$}&{$0.1583$}&{$0.1322(15)$}&{$0.0063(2)$}&{$0.0452(11)$}&{$0.4024(268)$}\\

\hline \hline

  \multicolumn{6}{c}{$\beta = 5.8$} \\
\hline
$lattice$&$\kappa$& $am_{\pi}$& $am_q$&{$aF_{\pi}$}& {$am_N$}\\
\hline
{$32^3\times64$}&{$0.1543$}&{$0.1612(9)$}&{$0.0137(3)$}&{$0.0420(9)$}&{$0.3649(39)$}\\
{$~~~~~,,$}&{$0.15445$}&{$0.1355(9)$}&{$0.0096(3)$}&{$0.0366(11)$}&{$0.3418(40)$}\\
{$~~~~~,,$}&{$0.15455$}&{$0.1212(11)$}&{$0.0074(3)$}&{$0.0344(15)$}&{$0.3308(63)$}\\
{$~~~~~,,$}&{$0.15462$}&{$0.1050(10)$}&{$0.0055(2)$}&{$0.0328(12)$}&{$0.3100(168)$}\\
{$~~~~~,,$}&{$0.15466$}&{$0.1017(16)$}&{$0.0051(3)$}&{$0.0306(16)$}&{$-$}\\
{$~~~~~,,$}&{$0.1547$}&{$0.0833(22)$}&{$0.0032(3)$}&{$0.0274(23)$}&{$0.2854(185)$}\\
\hline \hline
\end{tabular}
\end{center}
\caption{Lattice data for $am_{\pi}$,  $am_q$, $aF_{\pi}$ and $am_N$.   }
\label{table2}
\end{table}

 
\section{Determination of $r_0\over a$}
  \begin{figure}
\begin{center}
 \includegraphics[width=3.5in,clip]
{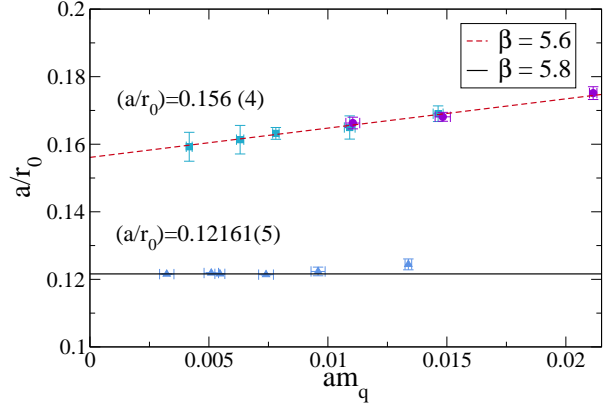}
\caption{$a/r_0$ versus $am_q$ at $\beta = 5.6$ and $\beta = 5.8$.}
\end{center}
\label{fig2}
\end{figure}
We measured the Wilson loops $\langle W(R,T)\rangle$ with temporal
extents up to $T=32~(24)$ and
spatial separations up to $R=\sqrt{3} \times 16~(\sqrt{3}\times 12)$ for lattice volume $32^3\times 64 ~(24^3\times 48)$.

A reasonable estimate of the static potential $aV(R)$ is obtained by
the plateau reached at large $T$ of the {\em effective} potential
\begin{equation}
aV_{\rm eff}(R,T) = 
{\rm ln}\frac{\langle W(R,T)\rangle}{\langle W(R,T+1)\rangle}. 
\end{equation}

Phenomenologically, the potential $V$ between
a static quark and an antiquark at a distance $r$ apart is parametrized as follows:
$V(r)\, =\, V_0\, + \,\sigma\,r\, +\, \frac{\alpha}{r} $
where $\sigma$ is the string tension which has the dimension of
mass$^2$.  In lattice units, we have
$a V(r) \,=\, a V_0 \,+\, a^2 \sigma\,\frac{r}{a} \,+\, {\alpha}\frac{a}{r}
$. Writing $ r=Ra$ and $\sigma = {\tilde \sigma}/a^2$, we get
$a V(R) \,=\, a V_0 \,+\, {\tilde \sigma} R\,+\, \frac{\alpha}{R}$.

Using the 
expression for the 
perturbative lattice Coulomb potential \cite{langrebbi, michael}
\be
\left[\frac{1}{R}\right]~=~ \frac{4 \pi}{L^3}~ \sum_{q_{i}\neq 0}~ 
\frac{{\rm cos} (a q_{i}\cdot R)}{4 {\rm sin}^2 (a q_{i}/2)},
\ee
the parametrization of the corrected potential on the lattice reads
\be  aV(R) ~=~ aV_0 ~+~ {\tilde \sigma}~R 
~-~ \frac{\alpha}{R}~ - ~ \delta_{\rm ROT}~ 
\left ( \Bigg[\frac{1}{R}\Bigg] - \frac{1}{R}\right )\label{corr-pot}
\ee
where $\delta_{\rm ROT}$ is the coefficient of the correction term.
The measured static potential is fit to the formula in Eq. (\ref{corr-pot})
which corrects the lattice data for the lattice artifacts in the Coulomb 
potential. The first three terms of Eq. (\ref{corr-pot}) now
gives the continuum potential (i.e., without lattice artifacts). The 
inverse of the Sommer parameter ($r_0$) in lattice units is calculated using 
\be
\frac{a}{r_0} = \sqrt{ \frac{\tilde{\sigma}}{ 1.65 -\alpha}}.
\label{abyrc}
\ee
In Fig. \ref{fig2} we have plotted $a\over r_0$ versus lattice quark mass at 
$\beta$=5.6 and 5.8. In the linear fit at $\beta$=5.6, we have included 
$\kappa$=0.15775, 0.158, 0.158125 in the lattice volume $24^3\times 48$ and 
$\kappa$=0.158, 0.15815, 15825, 0.1583 in the lattice volume $32^3\times 64$. 
Similarly at $\beta$=5.8 the linear fit is done with  $\kappa$=0.15445, 0.15455,
0.15462, 0.15466 and 0.1547 in lattice volume $32^3\times 64$. In the chiral 
limit, the values of $a\over r_0$ obtained are 0.156(4) and 0.12161(5) at 
$\beta$=5.6 and 5.8 respectively.
\section{Cutoff effects}
A major source of concern with the use of unimproved Wilson fermions is the 
potential presence ${\cal O}(a)$ cutoff effects in various observables. In this 
section we present the cutoff dependence of a variety of observables. 
\begin{figure}
\subfigure{
 \includegraphics[width=3.5in,clip]
{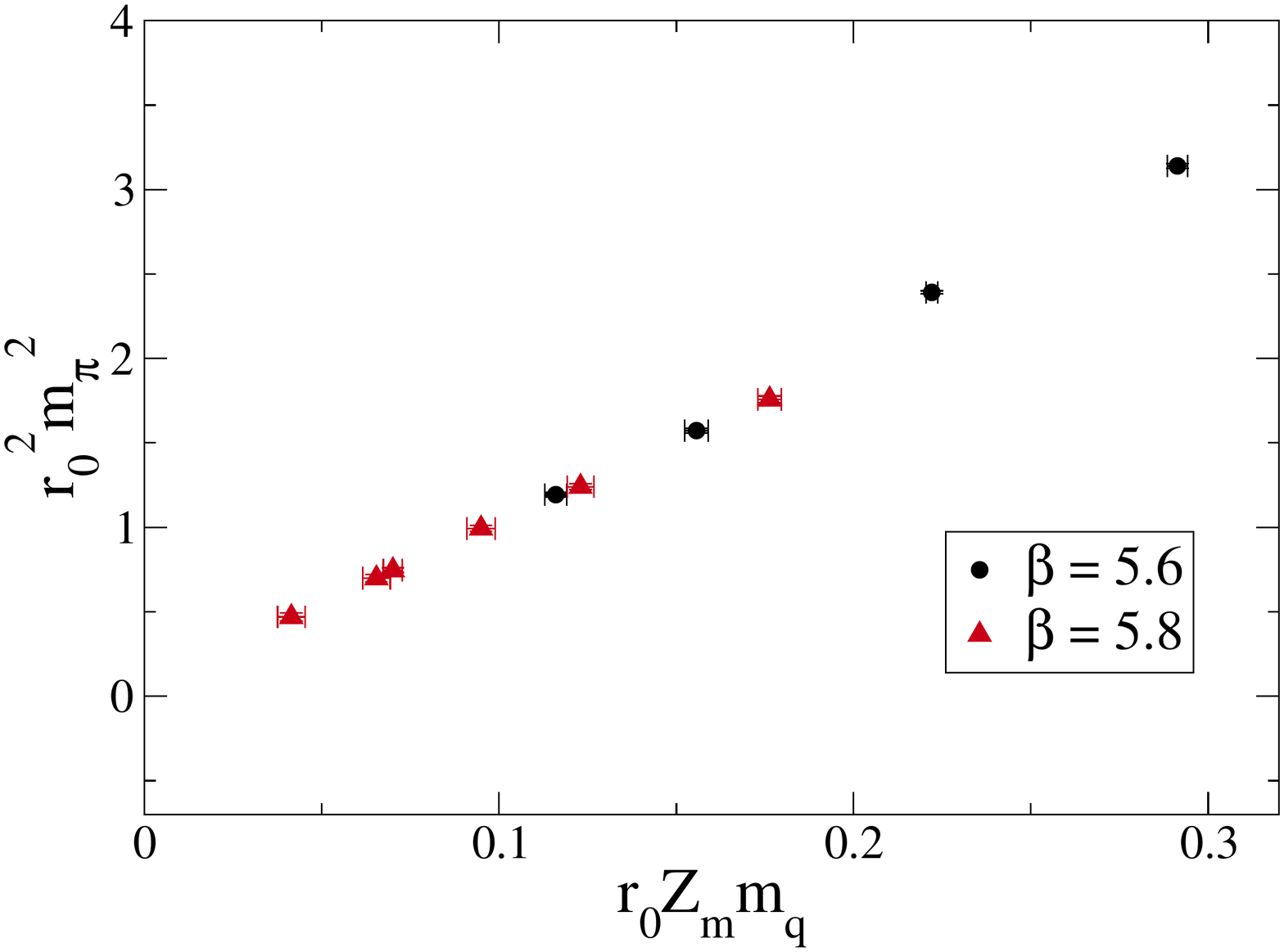}}
\subfigure{
 \includegraphics[width=3.5in,clip]
{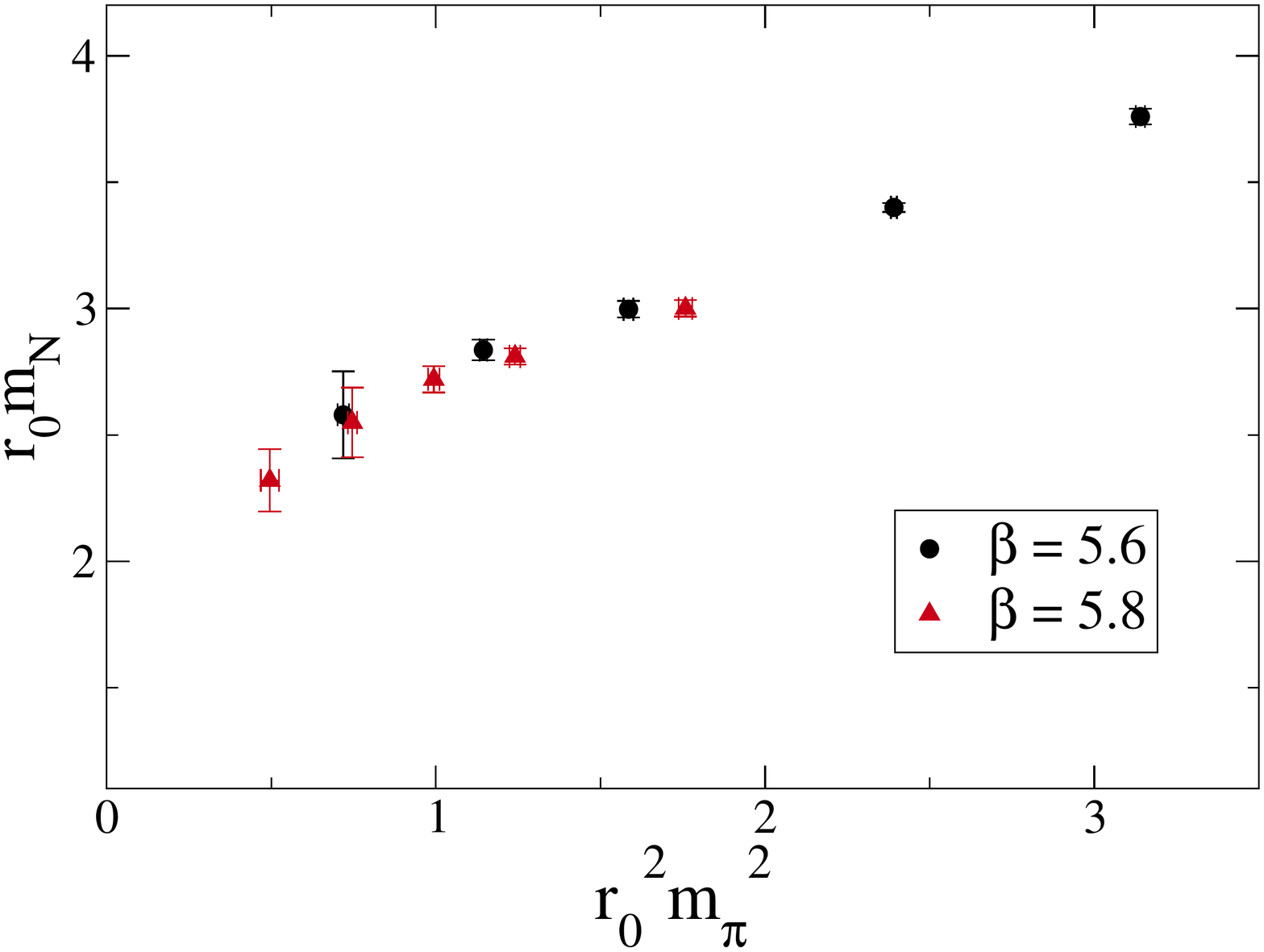}}
\caption{Scaling of pion mass squared (left) and nucleon
mass (right).}
\label{scaling}
\end{figure}
In Fig. \ref{scaling} we have plotted $(r_0m_\pi)^2$ (left figure) and $r_0 m_N$ 
(right figure) versus $r_0 Z_m m_q$ and $(r_0 m_\pi)^2$ respectively using the data
at $\beta$=5.6 and 5.8. Here $Z_m$ is the quark mass renormalization constant which we have 
calculated from the $Z$ factors given in \cite{becirevic}. 
It is seen that scaling violation are negligible.   

\begin{figure}
 \includegraphics[width=3.5in,clip]
{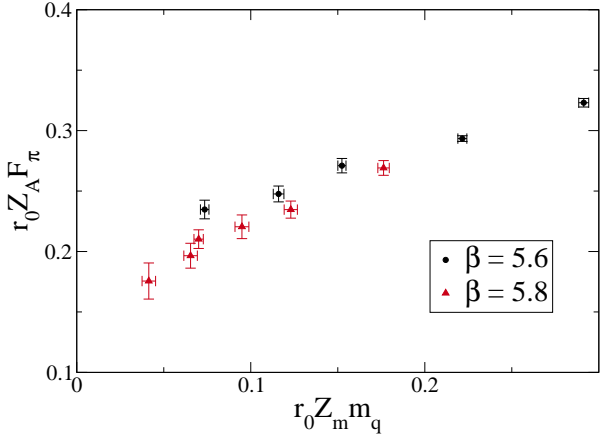}
\caption{Scaling of $F_\pi$.}
\label{fig8a}
\end{figure}

In Fig. \ref{fig8a} we have plotted $r_0 Z_A F_\pi$ versus  $r_0 Z_m m_q$ 
using the data at $\beta$=5.6 and 5.8. Data at $\beta$=5.6, except at the 
smallest quark mass, is from lattice volume $24^3\times 48$ and the smallest
quark mass data ($\kappa$=0.1583) is from lattice volume $32^3\times 64$. The
data at $\beta$=5.8 is from lattice volume $32^3\times 64$. Note that the 
physical volume at  $\beta$=5.6 and lattice volume $24^3\times 48$ is very
close to the physical volume at  $\beta$=5.8 and lattice 
volume $32^3\times 64$. 
From the figure it appears that $F_\pi$ data at $\beta$=5.6 and 5.8 do not
exhibit scaling for smaller quark mass region.
Since $F_\pi$ has larger finite volume effect than 
$m_\pi$, the deviation seen at small quark mass region could be partially due to
finite volume effects. Note that we have taken the values of non-perturbative 
renormalization constant $Z_A$ at $\beta$=5.6 and 5.8 for unimproved Wilson 
fermions from Ref. \cite{becirevic}. In this reference at $\beta$=5.8 the 
lattice volume was $24^3\times 48$ and the smallest quark mass probed was 
greater than $75$ MeV. Thus the systematic error in the chiral extrapolation of $Z_A$
performed in the Ref. \cite{becirevic} at $\beta$=5.8 may not be properly estimated.
Hence the apparent lack of scaling exhibited by $F_\pi$ may also be partially due to inaccurate 
determination of $Z_A$.

 \begin{figure}
\subfigure{
 \includegraphics[width=3.5in,clip]
{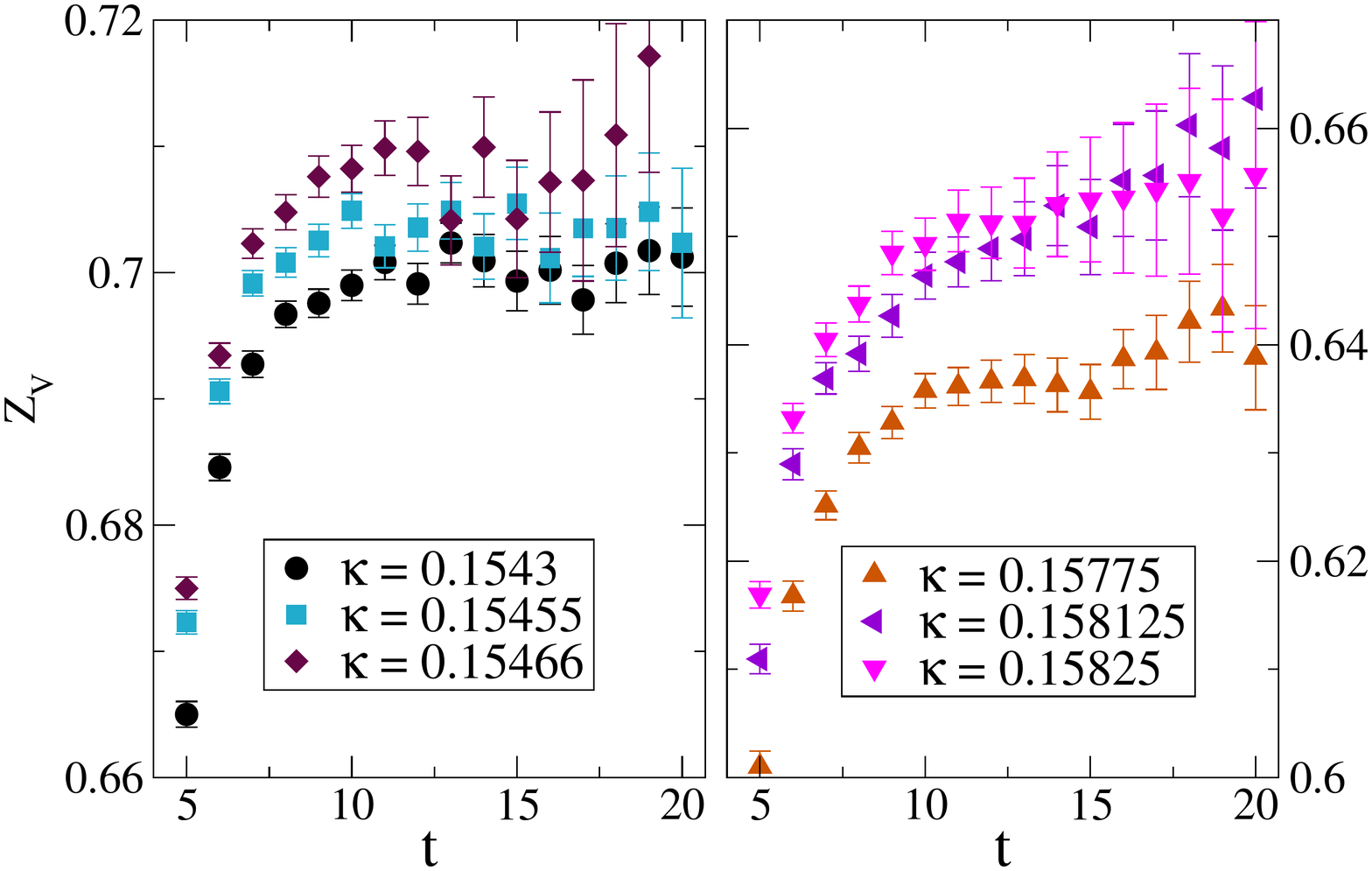}}
\subfigure{
 \includegraphics[width=3.5in,clip]
{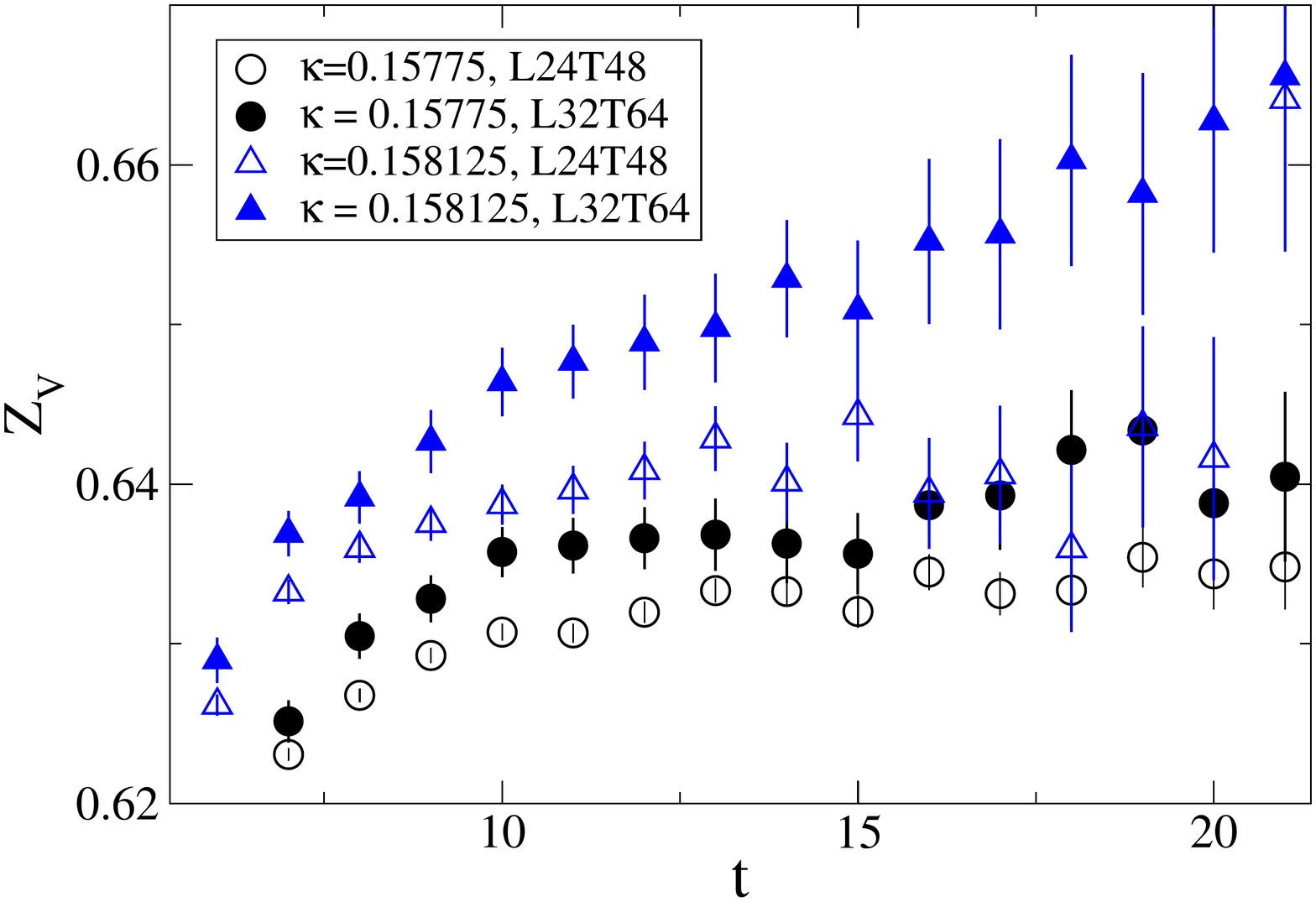}}
\caption{Quark mass and scale dependence of $Z_V$ at L32T64 (left), quark mass and volume dependence of $Z_V$
at $\beta = 5.6$ (right).}
\label{zfv}
\end{figure}

In the scaling region, hadron masses are expected to 
be independent of the lattice scale and our results demonstrate that they do 
exhibit very small, if not negligible, cut off dependence in the range of lattice spacings 
studied.

On the other hand the lattice renormalization constant $Z_V$ corresponding 
to the local vector current $V_\mu^{local}$ which is not conserved for $a\neq 0$ 
depends on the lattice spacing and is expected to approach unity as 
lattice spacing $a \rightarrow 0$. In order to verify the expected lattice scale dependence of $Z_V$
we calculate $Z_V$ from 
\cite{maiani}
\begin{eqnarray}
R_V(t) = \frac{\sum_{k=1}^{3} \sum_x \langle {\hat V}_k(x,t) 
V^{\rm local \dagger }_k (x,t) \rangle }
{\sum_{k=1}^{3} \sum_x \langle V^{\rm local} _k(x,t) 
V^{\rm local \dagger}_k (x,t) \rangle } = Z_V ~+~ \ldots
\end{eqnarray}
with 
\begin{eqnarray}
V^{\rm local}_\mu (x) = {\bar q}_i(x) \gamma_\mu q_j(x)
\end{eqnarray}
and (using r=1 throughout in this work)
\begin{eqnarray}
{\hat V}_\mu(x) =  \frac{1}{2} \left [ {\bar q}_i(x)  (\gamma_\mu -1 )
U_\mu(x) q_j(x+\mu)~ + ~
{\bar q}_i(x + \mu) (\gamma_\mu +1) U_{\mu}^\dagger(x) q_j(x) \right ]~.
\end{eqnarray}
 Our result shown in 
Fig. \ref{zfv} (left) exhibits the expected behaviour. On the other hand, finite 
volume effect on $Z_V$ is small though not negligible as shown in 
Fig. \ref{zfv} (right). As already discussed, in order to examine the cutoff effects 
in pion decay constant we need accurate determination of $Z_A$ in the chiral region which
is beyond the scope of the present paper.    

\section{Scale determination, chiral behaviour of nucleon mass and extraction of sigma term}

\begin{figure}
\subfigure{
 \includegraphics[width=3.5in,clip]
{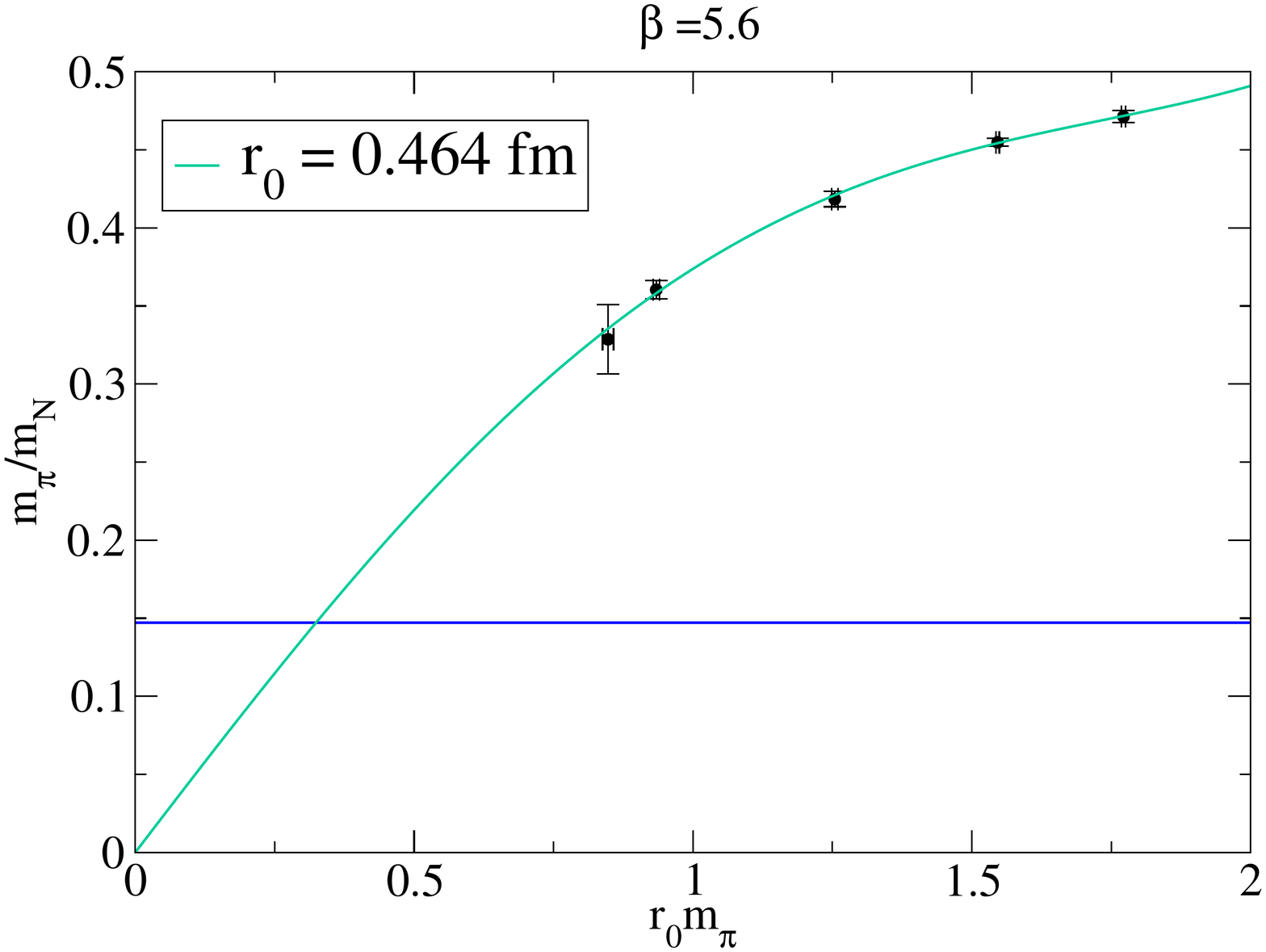}}
\subfigure{
 \includegraphics[width=3.5in,clip]
{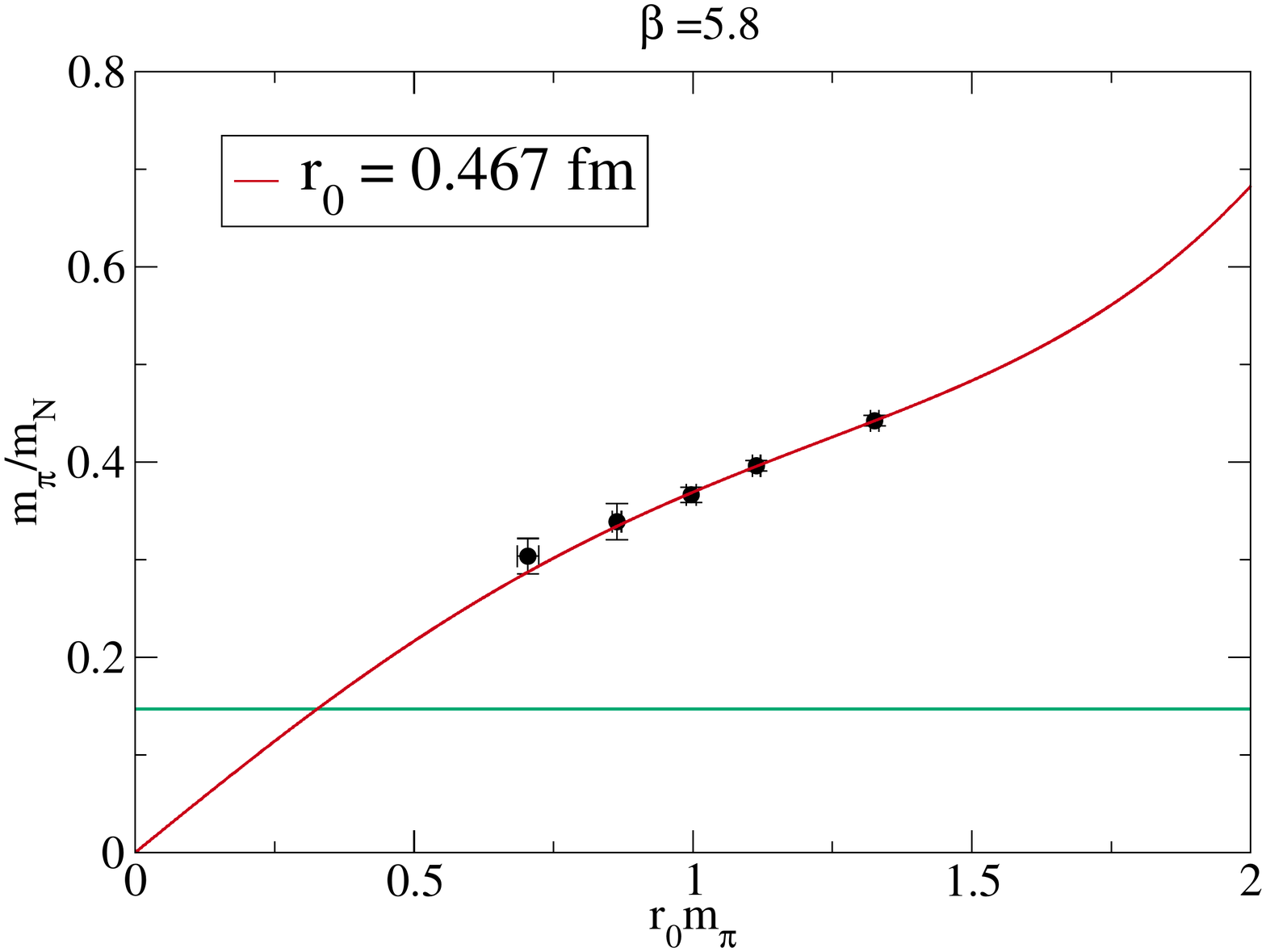}}
\caption{$m_\pi/m_N$ versus $r_0 m_\pi$ at $\beta$=5.6 (left) and $\beta$=5.8 (right) }
\label{mNvsmpisq}
\end{figure}

To extract the lattice scale, in Fig. \ref{mNvsmpisq} we have plotted the ratio $m_\pi\over m_N$ versus
$r_0 m_\pi$. The data is used to fit the phenomenological formula
\beq
\frac {m_\pi}{ m_N} = a_1 r_0 m_\pi -a_2(r_0 m_\pi)^3+ a_3 (r_0 m_\pi)^4
\eeq
motivated by baryon chiral perturbation theory. We fit the data at $\beta$=5.8,
lattice volume $32^3 \times 64$ and $\kappa$= 0.1543, 0.15445, 0.15455 
 and 0.15462. At $\beta$=5.6, we use $\kappa$= 0.1575, 0.15775 in lattice 
volume $24^3 \times 48$ and $\kappa$= 0.158, 0.15825, 0.1583 at lattice 
volume $32^3 \times 64$. Using the physical value of  $m_\pi\over m_N$, we 
get the intercept of the fitting curve and from the intercept we extract the 
pion mass in unit of $r_0$ at the physical point. Now using the value of physical pion
mass and the chiral limit of $\frac{a}{r_0}$, we have computed lattice spacing $a$. 
The computed values of lattice spacings at $\beta =$5.6 and 5.8 are 
$0.072$ fm and $0.057$ fm respectively.

\begin{figure}
 \includegraphics[width=3.5in,clip]
{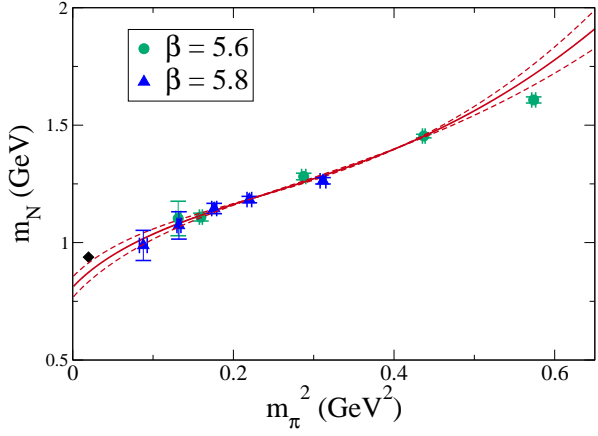}
\caption{Nucleon mass versus $m_\pi^2$ at $\beta$=5.6 and 5.8.}
\label{nucleonmass}
\end{figure}

We have plotted dimensionfull nucleon mass versus dimensionfull $m_\pi^2$ in
Fig. \ref{nucleonmass}. At $\beta$=5.6 we have plotted $\kappa$=0.1575, 0.15775
at lattice volume $24^3\times48$ and $\kappa$= 0.158, 0.15825, 0.1583 in 
lattice volume $32^3\times64$. At $\beta$=5.8 we have plotted $\kappa$= 0.1543,
0.15445, 0.15455, 0.15462 and 0.1547 in lattice volume $32^3\times 64$. To fit
the data we have excluded $\kappa$= 0.1575 and 0.1547 at $\beta$= 5.6 and 
5.8 respectively.
The data at $\kappa$=0.1575 is dropped because it is beyond the chiral regime 
and data at $\kappa$= 0.15475 has not negligible finite volume effect.     
We fit the data using  baryon chiral perturbation theory formula to 
order ($m_\pi^4$) given
in Ref. \cite{bali}:
\begin{eqnarray}
M_N~ &=&~ M_0 - 4 c_1~m_\pi^2 - \frac{3g_A^2}{32 \pi F_\pi^2}~m_\pi^3 
 ~+~ 4e_1^r~m_\pi^4 \nonumber \\
&~~~~&+ ~\frac{m_\pi^4}{8 \pi^2 F_\pi^2}~ \Bigg [ 
\frac{3c_2}{16}~ - ~ \frac{3g_A^2}{8 M_0}~ + ~ {\rm log} \frac{m_\pi}{\lambda}~
\Big ( 8c_1~ ~ - ~ \frac{3 c_2}{4}~ -~ 3 c_3~ - \frac{3g_A^2}{4M_0} \Big )
\Bigg ]~.
\end{eqnarray} 
For the fit we have treated $M_0$, $c_1$ and $e_1^r$ as free parameters and set 
$F_\pi$= 0.086 GeV, $g_A$= 1.256, $c_2$= 3.3 ${\rm GeV}^{-1}$, $c_3$= -4.7
${\rm GeV}^{-1}$ and $\lambda$= 1 GeV. Note that the values of $c_2$ and $c_3$
chosen are close to their phenomenological values \cite{bali}.  
The fit along with the error is shown in Fig. \ref{nucleonmass}.
The values of the parameters $M_0$, $c_1$ and $e_1^r$ we obtain from the fitting are $M_0 = 0.81(4)~{\rm GeV}$,
$c_1=-1.04(5)~{\rm GeV}^{-1}$ and $e_1^r=1.2(1)~{\rm GeV}^{-3}$.
Note that the $c_1$ obtained is close to the phenomenologically determined value \cite{bernard}. 

\begin{figure}
 \includegraphics[width=3.5in,clip]
{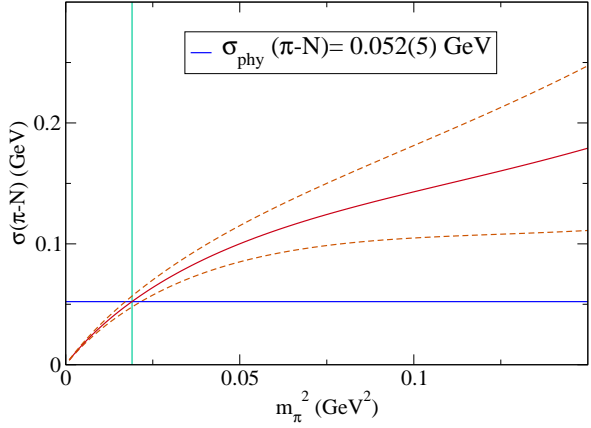}
\caption{Behaviour of the sigma term versus $m_\pi^2$ at $\beta$=5.6 and 5.8.}
\label{sigmaterm}
\end{figure}

Both from experimental and theoretical point of view, the pion-nucleon $\sigma$
term is of significant interest. A direct calculation of $\sigma$ involves 
the computation of a three point function and is beyond the scope of the 
present paper. An alternative method of calculation employs Feynman-Hellmann
theorem utilizing the dependence of nucleon mass on the quark mass which in 
tern can be converted into the dependence on pion mass squared. 
With  the parameters used to fit the nucleon data, we calculate  the sigma term
using the expression \cite{bali}
\begin{eqnarray}
\sigma~ &=&~  - 4 c_1~m_\pi^2 - \frac{9g_A^2}{64 \pi F_\pi^2}~m_\pi^3 
~+ ~m_\pi^4~ \Bigg [ 8 e_1^r - \frac{8 c_1 l_3^r}{F_\pi^2}~+~ 
\frac{3c_1}{8 \pi^2 F_\pi^2}~- ~ \frac{3c_3}{16 \pi^2 F_\pi^2} \nonumber \\ 
&~~~~&- \frac{9g_A^2}{64\pi^2M_0 F_\pi^2} ~+~ \frac{1}{4 \pi^2 F_\pi^2}
 ~ {\rm log} \frac{m_\pi}{\lambda}~
\Big ( 7c_1~ ~ - ~ \frac{3 c_2}{4}~ -~ 3 c_3~ - \frac{3g_A^2}{4M_0} \Big )
\Bigg ]~. \label{sigma}
\end{eqnarray}
where $l_3^r \equiv -\frac{1}{64\pi^2}\left({\bar l}_3 +2~{\rm log}~
 \frac{m_\pi^{phys}}{\lambda}\right)$.
In Fig. \ref{sigmaterm}, we plot $\sigma$ given in Eq. (\ref{sigma}) together with the 
error as a function of $m_\pi^2$ and at physical point $\sigma$ = 0.052(4) GeV
which is compatible with the currently available determinations \cite{bali}.

\section{Chiral extrapolations of pion mass and decay constant and quark condensate}
In this section we discuss chiral extrapolation of our data obtained from the pion propagators. 
\begin{figure}
 \includegraphics[width=3.5in,clip]
{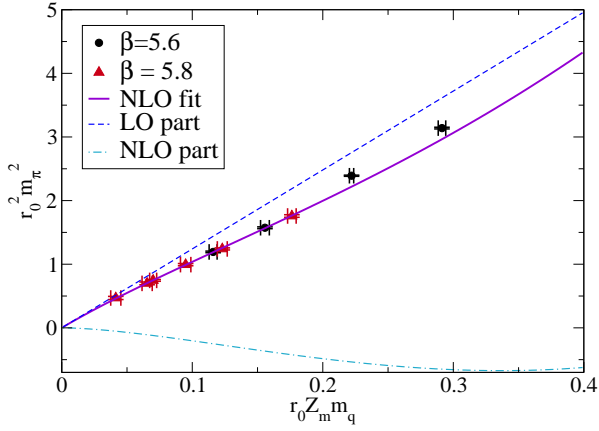}
\caption{NLO chiral perturbation theory fit to $(r_0 m_\pi)^2$ versus 
$r_0 Z_m m_q$.}
\label{fig6}
\end{figure}
In SU(2) $\chi$PT at NLO \cite{gl}, the quark mass ($r_0 m_q$) dependence of 
$( r_0 m_\pi)^2$ is given by 
\be
(r_0 m_\pi)^2 &=& 2r_0^2m_qB \left[ 1-\frac{m_qB}{16\pi^2F^2}{\rm ln}
\frac{\Lambda_3^2}{2m_qB} \right]
\ee
where $F$ is the chiral limit of the pion decay constant and $B$ 
and $\Lambda_3$ are low energy constants. 
In Fig. \ref{fig6}
we compare our data at $\beta=5.6$ and $5.8$ with a fit to the NLO $\chi$PT formula 
(solid line). For clarity, the LO (dashed line)
and the NLO (dot-dashed line) contributions are separately 
shown. At $\beta=5.6$ and lattice volume $24^3 \times 48 $, $\kappa$= 0.158 and
0.158125 and at $\beta=5.8$ and lattice volume $32^3 \times 64 $, 
$\kappa$= 0.1543, 0.15445, 0.15455, 0.15462 and 0.15466 are used for 
fitting. The value of $F=86~{\rm MeV}$ is taken as an input. 
The values of the parameters $r_0B$ and $r_0\Lambda_3$ obtained from the fit are
$r_0B=6.20(13)$ and $r_0\Lambda_3=2.62(17)$. Converting these values in physical unit
we get $B=2651(56)$ MeV and $\Lambda_3=1120(73)$ MeV. Note that the currently
available estimates of $B$ and $\Lambda_3$ are $2112\lesssim B \lesssim 2811$ MeV and
 $458\lesssim \Lambda_3 \lesssim 1020$ MeV \cite{flag}. Thus we find that the quark mass dependence of pion mass
squared exhibited by our data is in accordance with NLO $\chi$PT.

To expose the presence of the chiral logarithm, it is customary to plot
\be
\frac{(r_0 m_\pi)^2}{r_0 m_q} &=& 2 r_0 B \left[ 1 - \frac{m_qB}{16\pi^2F^2}
{\rm ln}
\frac{\Lambda_3^2}{2m_qB} \right] 
\ee
\begin{figure}
 \includegraphics[width=3.5in,clip]
{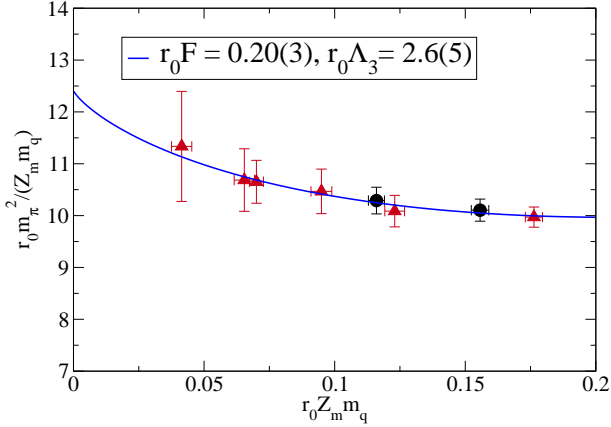}
\caption{$\frac{r_0m_\pi^2}{Z_mm_q}$ versus $r_0Z_mm_q$ for $\beta=5.6$ and $\beta = 5.8$.}
\label{fig7}
\end{figure}
We plot $ \frac{(r_0m_\pi)^2}{r_0 m_q}$ versus $r_0 Z_m  m_q$ in 
Fig. \ref{fig7}. In the fit, we have used $r_0B = 6.2029$ which is obtained 
from the  fit in Fig. \ref{fig6}. The quark masses used in the fit are the 
same as in Fig. \ref{fig6}.

\begin{figure}
 \includegraphics[width=3.5in,clip]
{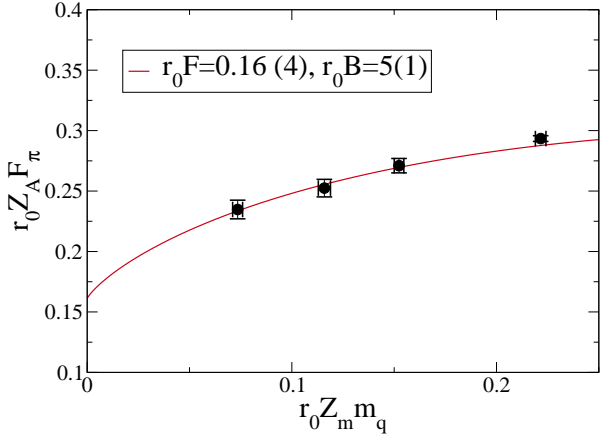}
\caption{Chiral extrapolation of $r_0 F_\pi$ at $\beta$=5.6.}
\label{fig8}
\end{figure}

The quark mass dependence of the pion decay constant $r_0F_\pi$ in NLO
$\chi$PT is given by 
\be
r_0F_\pi &=& r_0F \left[ 1+\frac{m_qB}{8\pi^2F^2}{\rm ln}
\frac{\Lambda_4^2}{2m_qB} \right]
\ee
where $\Lambda_4$ is another low energy constant. 
As already mentioned, our renormalized $F_\pi$ data at $\beta=5.8$ has larger uncertainty
coming from $Z_A$ determination and also possible finite volume effects at the smaller 
quark masses. Therefore we use the data only at $\beta=5.6$ for our exploratory
chiral extrapolation.
In Fig. \ref{fig8} we 
compare our data for $r_0F_\pi$ versus $r_0 Z_m m_q$ with NLO $\chi$PT at 
$\beta$=5.6. 
Since $F_\pi$ has larger finite volume effect compared to $m_\pi$,
we have used data in lattice volume $32^3 \times 64$ in this fit. Appropriate 
to the chiral region, we have used only the lower quark masses 
($\kappa$=0.158, 0.15815 and 0.1583) for the fit. Since we have fewer 
points to fit, we have used $r_0 \Lambda_4$ = 3.1134 ($\Lambda_4=1324$ MeV) (see \cite{flag}) as the input 
to the fit. The upper limits of the values obtained for $F$ and $B$ (see Fig. \ref{fig8}) 
are close to the currently accepted values \cite{flag}. Note however that there is some uncertainty
arising from the value of $Z_A$.
Thus in the case of $F_\pi$ it is safe to say that our data is not incompatible with
NLO $\chi$PT prediction. However simulations at smaller quark masses with less systematic errors due to finite volume and $Z_A$ 
determination
is needed to reach a definite conclusion.

Next we consider the extraction of chiral condensate from our data. 
In continuum limit, the effect of Wilson term  on
chiral condensate does not vanish because Wilson term is a dimension five 
operator. In other words, it renormalizes the chiral condensate additively, in
addition to multiplicatively. Thus in the case of Wilson fermions, a  direct measurement of the quark 
condensate on the lattice and then taking the chiral continuum limit will 
not give the desired condensate. The well-known alternate procedure to 
calculate the chiral condensate on the lattice with Wilson fermions is
utilizing chiral Ward-Takahashi identities \cite{bochicchio,giusti} 	

\begin{figure}
 \includegraphics[width=3.5in,clip]
{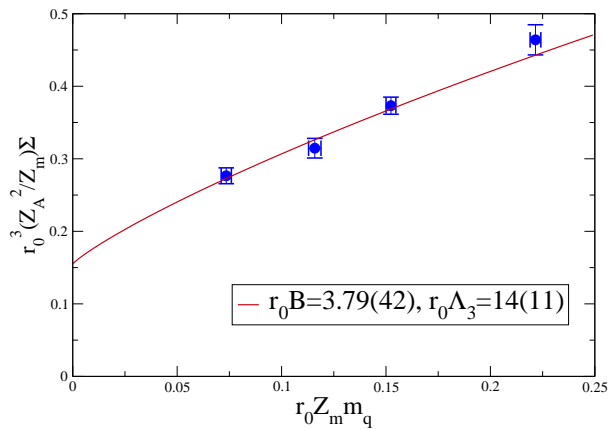}
\caption{Renormalized quark condensate versus renormalized PCAC quark mass at $\beta = 5.6$ in unit of $r_0$.}
\label{fig9}
\end{figure}

\be
\langle {\overline \psi} \psi \rangle ~ \delta^{ab}= 2~ m_q \int d^4x ~
\langle P^a(x)P^b(0)\rangle~=> \langle {\overline \psi} \psi \rangle~= ~2 m_q \frac{C_{PP}}{m_\pi} ~ = ~
C_{AP} 
\ee
using Eq. (\ref{mqap}).  In Fig. \ref{fig9} we plot the chiral behaviour of 
the quark condensate ($\Sigma$) together with an NLO fit at $\beta$=5.6 and lattice volume
$32^3 \times 64$.
According to NLO $\chi$PT 
\be
\Sigma=F^2 B\left[1-\frac{3Bm_q}{16\pi^2F^2}\ln(\frac{2Bm_q}{\Lambda_3^2})
\right]
\ee
The quark masses used are as in Fig. \ref{fig8}. In this fit 
we have used the chiral limit of $r_0 F$ =0.20 ( corresponds to $F=$ 86 MeV) as input.The 
value of chiral condensate obtained from the fit is $\Sigma^{1/3}$ = 228(8) MeV.
Note that since the values of $F_\pi$ have entered the determinations of condensates
all the caveats that we have mentioned in the context of chiral extrapolation of 
$F_\pi$ apply in this case also.

\section{Summary}
We have calculated pion mass, pion decay constant, PCAC quark mass and nucleon mass
in two flavour lattice QCD with unimproved Wilson fermion and gauge actions. 
Simulations are performed using DD-HMC algorithm at two lattice spacings and two volumes 
for several values of the quark mass. The cutoff effects in pion mass and nucleon mass
for the explored region of parameter space are found to be negligible. The chiral behaviours of 
pion mass, pion decay constant and quark condensate are found to be qualitatively consistent with
NLO chiral perturbation theory.
\vskip .25in
{\bf Acknowledgements}
\vskip .1in
  Numerical calculations are carried out on Cray XD1 and Cray XT5 systems 
supported 
by the 10th - 12th Five Year Plan Projects of the Theory Division, SINP under
the DAE, Govt. of India. We thank Richard Chang for the prompt maintainance of 
the systems and the help in data management. This work was in part based on 
the public lattice gauge theory codes of the 
MILC collaboration \cite{milc} and  Martin L\"{u}scher \cite{ddhmc}.



\end{document}